\documentclass[aip,reprint]{revtex4-1}
\usepackage{graphicx}
\usepackage{siunitx}     % for units like \SI{3}{\micro\meter}
\usepackage{amsmath}
\usepackage{color}
\usepackage{orcidlink}

\begin{document}

\title{FinLN: 3D Fin Lithium Niobate Acoustic Resonators}

\author{Haorui Ni\,\orcidlink{0009-0000-7966-0842}}
\email{nih@purdue.edu}
\author{Sunil A. Bhave\orcidlink{0000-0001-7193-2241}}
\affiliation{OxideMEMS Laboratory, Purdue University, West Lafayette, IN 47907, USA}

\begin{abstract}
We demonstrate a three-dimensional fin lithium niobate (FinLN) acoustic resonator fabricated from thick lithium niobate on insulator using deep ion etching and sidewall electrode patterning. The FinLN geometry enables strong three-dimensional acoustic confinement and enhanced electromechanical coupling while maintaining a compact device footprint. The fabricated devices operate near 300~MHz and exhibit an effective electromechanical coupling coefficient of 6.2\% with a mechanical quality factor of 430, in good agreement with finite-element simulations. Compared to planar surface acoustic wave resonators fabricated on the same wafer, the FinLN configuration achieves an 1.8$\times$ enhancement in effective electromechanical coupling. This work establishes FinLN as a promising platform for low-cost, compact and high-performance RF MEMS and piezo-optomechanical systems.
\end{abstract}

\maketitle

Microacoustic devices operating in the hundreds of MHz to GHz frequency range are key components for sensing\cite{MandalSurface2022,LiFrequency2023a}, signal processing\cite{luRecentAdvancesHighperformance2025,YangSVSAW2025,GiribaldiCompact2024}, and emerging quantum technologies\cite{SatzingerQuantum2018,JiangEfficient2020,WeaverIntegrated2024,wendt2026electrically,HugotApproaching2026}. Owing to strong electromechanical interaction and compact wavelength scaling, acoustic platforms enable miniaturized resonators and filters that are difficult to realize using electromagnetic approaches. Continued scaling of these systems simultaneously demands higher electromechanical coupling and smaller device footprint for dense integration and low-loss operation.

Transferred thin-film lithium niobate (LN) has emerged as a leading piezoelectric platform for integrated acoustic devices, providing single-crystal material quality with superior piezoelectric coefficients and low acoustic loss, while enabling resonance frequency definition through lithographic control of the interdigital transducer (IDT) pitch. High-performance LN acoustic resonators have been demonstrated using suspended and thickness-trimmed thin-film LN membranes\cite{anusornPracticalDemonstrationsFR3Band2025,WangDesign2015}, periodically poled piezoelectric film (P3F) LN\cite{kramerAcousticResonators1002025}, and thin-film LN bonded onto high phase velocity substrates such as sapphire or SiC\cite{zhang2020surface}. Despite these advances, current thin-film LN platforms with LN film thicknesses below $\sim$1~$\mu$m rely on ion-slicing technologies such as Smart-Cut\cite{ButaudSmart2020a}, which remain costly, process-intensive, and limited in wafer availability. In contrast, thick piezoelectric films ($>$5~$\mu$m)\cite{ParkPreparation2010}, including lithium tantalate bonded to silicon using bonding and lapping processes, offer significantly lower cost and repeatable manufacturability, and have recently enabled acoustic filter implementations\cite{rubyNovelTemperatureCompensatedSilicon2021}. Motivated by these developments, we adopt a similar strategy using thick lithium niobate as the device layer, leveraging its excellent piezoelectric properties while avoiding the complexity associated with ion-sliced thin films.

In this work, we introduce a three-dimensional fin-based lithium niobate (FinLN) acoustic resonator fabricated from thick LN, inspired by the transition from planar MOSFETs to FinFET architectures for enhanced electrostatic control and scaling\cite{HisamotoFinFETa2000}. The demonstrated devices operate at 300~MHz and exhibit an effective electromechanical coupling coefficient of 6.2\%, while maintaining a compact footprint enabled by deep Ar ion etching and conformal sidewall metal deposition and patterning.

\begin{figure}[t]
\centering
\includegraphics[width=\columnwidth]{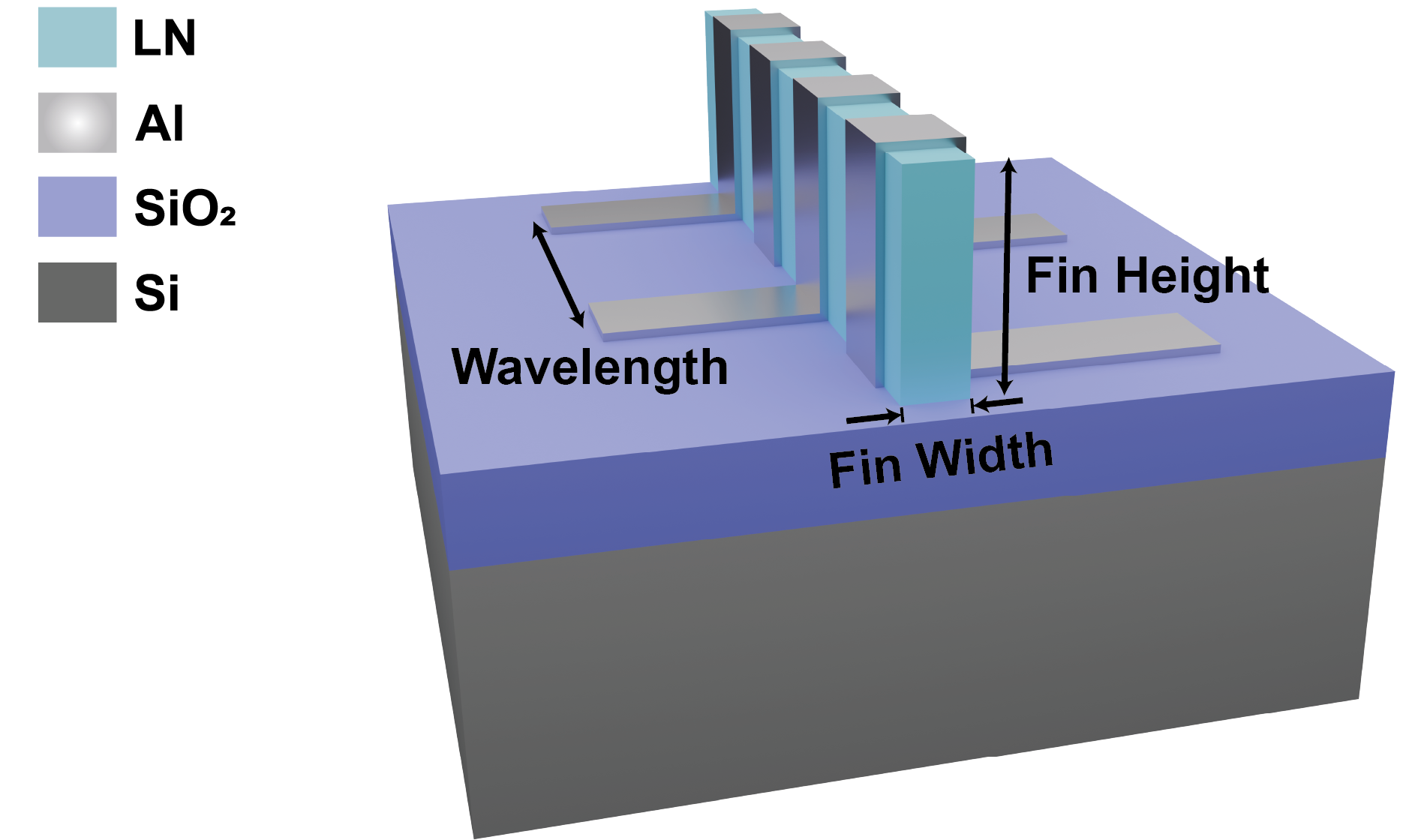}
\caption{Schematic illustration of the fin lithium niobate (FinLN) acoustic resonator. Two unit cells are shown, each containing one pair of interdigital transducer (IDT) electrodes patterned across the fin. The fin width, fin height, and acoustic wavelength are indicated.}
\label{fig:1}
\end{figure}

\begin{figure}[t]
\centering
\includegraphics[width=\columnwidth]{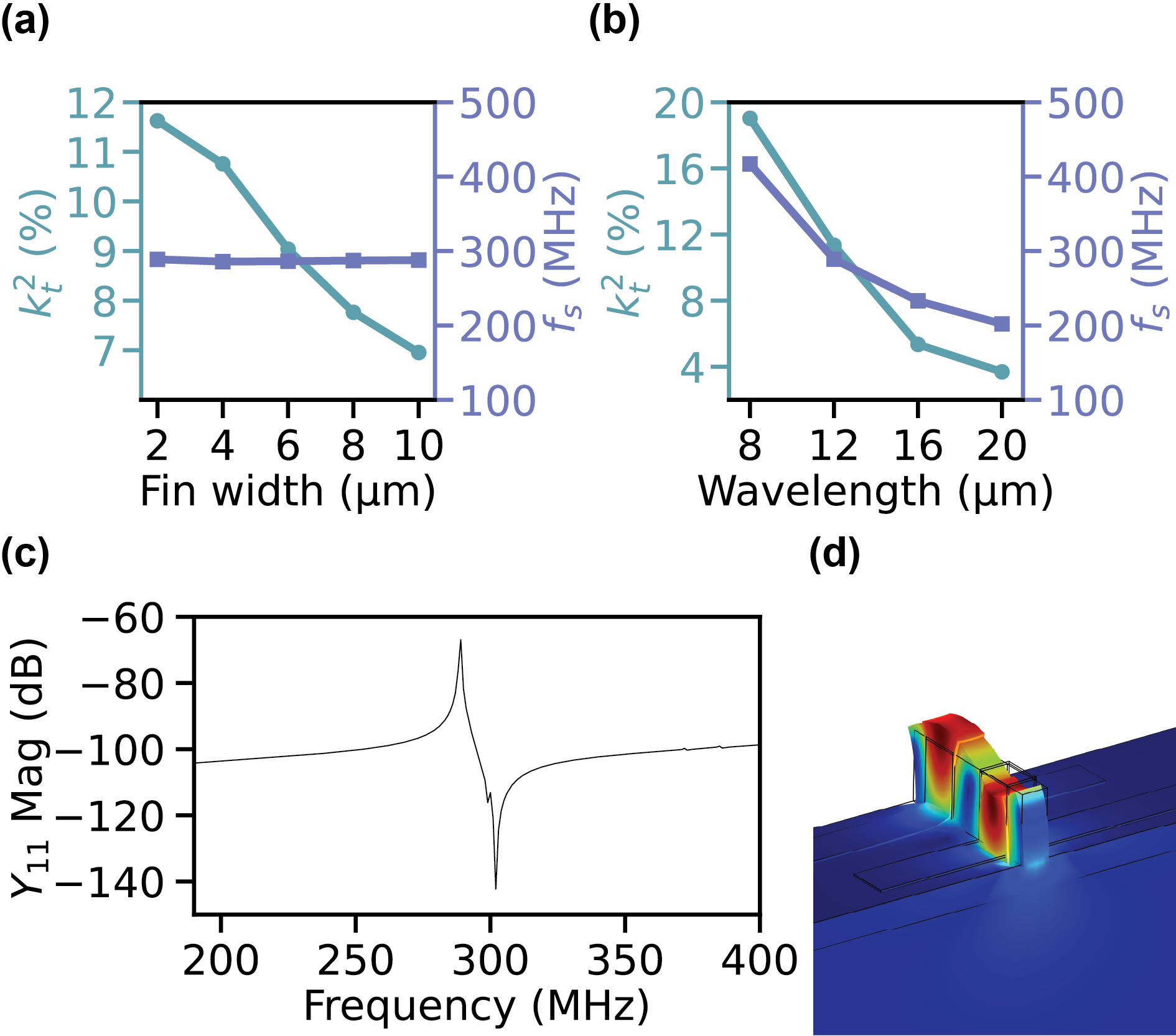}
\caption{Finite-element simulation results of the unit cell of the FinLN acoustic resonator.
(a) Electromechanical coupling coefficient $k_t^2$ and resonance frequency $f_s$ as a function of fin width with the acoustic wavelength fixed at 12~$\mu$m.
(b) $k_t^2$ and $f_s$ as a function of acoustic wavelength with the fin width fixed at 2~$\mu$m.
(c) Electrical admittance magnitude with the acoustic wavelength fixed at 12~$\mu$m and the fin width fixed at 2~$\mu$m.
(d) Resonance mode shape of unit cell at 290~MHz using the same parameters as in (c).}
\label{fig:2}
\end{figure}

The proposed FinLN acoustic resonator consists of a high-aspect-ratio LN fin fabricated on a 2~$\mu$m SiO$_2$ layer supported by a Si substrate (Fig.~\ref{fig:1}). Two identical unit cells of the FinLN resonator are shown, each containing one pair of aluminum interdigital transducer (IDT) electrodes patterned on the sidewalls in a ground--signal configuration. This IDT configuration enables electrical excitation of confined acoustic modes within the FinLN waveguide. Similar to conventional surface acoustic wave and lamb wave transducers, the alternating electrode polarity generates a periodic electric field that couples to the piezoelectric FinLN and launches confined acoustic vibration along the fin length. Distinct from planar acoustic devices, the FinLN geometry leverages a high-aspect-ratio fin to expand the functional acoustic volume while simultaneously confining the acoustic mode, analogous to the transition from planar MOSFETs to non-planar FinFET architectures, and from silicon membrane resonators to comb-drive resonators\cite{tang1990electrostatic}, resonant body transistors\cite{weinstein2010resonant}, and silicon bars\cite{hakim2024ferroelectric}. The electric field applied through the sidewall electrodes in the FinLN resonator is primarily oriented along the fin length direction, which suppresses fringing fields and enhances spatial overlap between the electric field and the confined acoustic mode. In conventional surface acoustic wave (SAW) devices, suppression of transverse modes typically requires an interdigital transducer (IDT) aperture of at least 20$\times$ the acoustic wavelength, resulting in a large unit-cell area defined by the product of aperture and wavelength. In contrast, the FinLN resonator does not rely on a wide aperture for transverse mode suppression. Instead, unwanted transverse modes are intrinsically suppressed by the high-aspect-ratio fin cavity, which provides strong lateral acoustic confinement. As a result, the FinLN architecture achieves enhanced electromechanical coupling while maintaining a compact device footprint through three-dimensional geometric confinement.

To identify appropriate design parameters for the FinLN resonator, three-dimensional finite-element simulations of the single periodic unit cell were performed using COMSOL Multiphysics (Fig.~\ref{fig:2}). To remain consistent with fabrication capability, the fin height was fixed at 5~$\mu$m. The acoustic wavelength and fin width were then systematically varied to evaluate the electromechanical coupling coefficient $k_t^2$ and resonance frequency $f_s$. Fig.~\ref{fig:2}(a) shows $k_t^2$ and $f_s$ as a function of fin width with the acoustic wavelength fixed at 12~$\mu$m, while Fig.~\ref{fig:2}(b) presents $k_t^2$ and $f_s$ as a function of acoustic wavelength with the fin width fixed at 2~$\mu$m. The simulation results indicate that a higher fin aspect ratio (defined as fin height-to-width ratio) and a shorter acoustic wavelength lead to enhanced electromechanical coupling, as $k_t^2$ exhibits a monotonic dependence on acoustic wavelength over the simulated parameter range. In contrast, the resonance frequency $f_s$ is relatively insensitive to fin width and increases with decreasing acoustic wavelength, consistent with conventional acoustic scaling behavior. Taking into account these trends and fabrication capabilities, a fin width of 2~$\mu$m and an acoustic wavelength of 12~$\mu$m were selected, after which a frequency-domain harmonic analysis of a single unit cell was performed to extract the electrical admittance response. Fig.~\ref{fig:2}(c) shows the simulated $Y_{11}$ admittance magnitude, and Fig.~\ref{fig:2}(d) illustrates the corresponding resonance mode shape of one unit cell of the FinLN resonator at approximately 290~MHz.

\begin{figure}[t]
\centering
\includegraphics[width=\columnwidth]{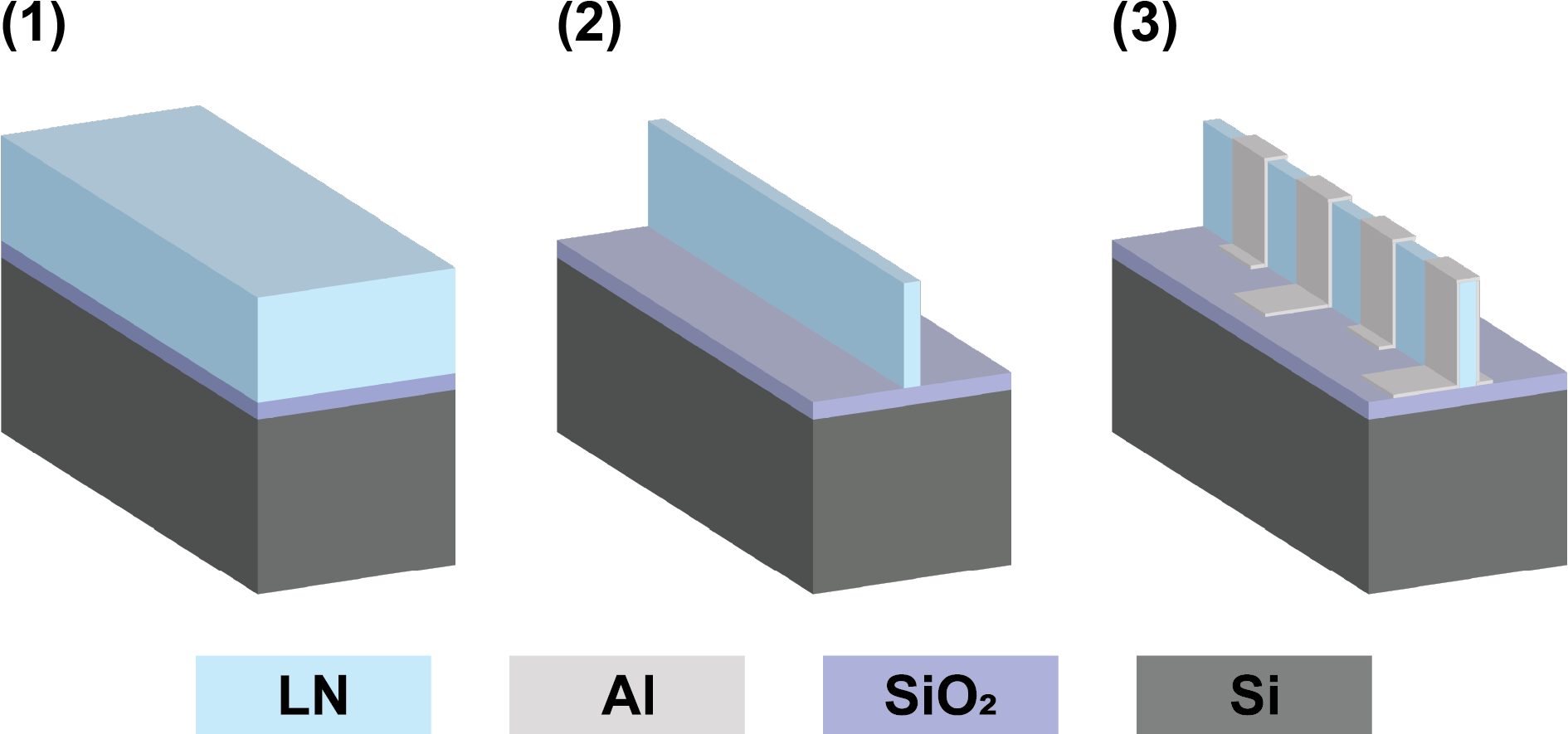}
\caption{Fabrication process flow of the FinLN devices: (1) starting from a thick Z-cut LNOI wafer with a 5~$\mu$m LN device layer, (2) Ar ion etching to define the fin structure, and (3) sidewall electrode formation by e-beam evaporation using glancing-angle deposition (GLAD), depositing 15~nm Ti and 200~nm Al.}
\label{fig:3}
\end{figure}

\begin{figure*}[t]
\centering
\includegraphics[width=\textwidth]{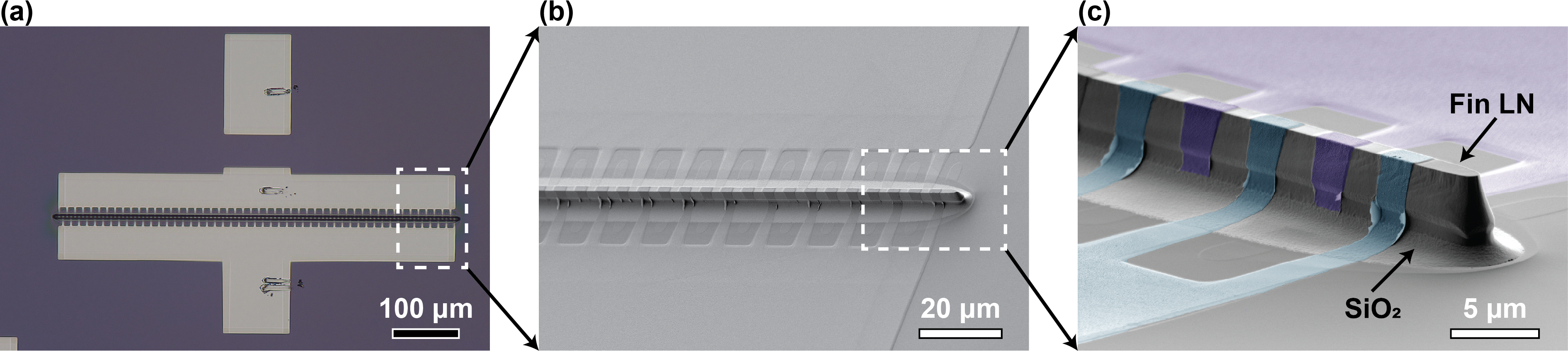}
\caption{Images of the fabricated fin lithium niobate (FinLN) acoustic resonator.
(a) Optical microscope image showing a top-view overview of the device.
(b) Scanning electron microscope (SEM) image of the top view.
(c) False-colored zoom-in SEM image of the device side view, where light blue indicates the ground electrode and light purple indicates the signal electrode.}
\label{fig:4}
\end{figure*}
The FinLN devices were fabricated on a Z-cut lithium niobate on insulator (LNOI) wafer consisting of a 5~$\mu$m LN device layer, a 2~$\mu$m buried SiO$_2$ layer, and a 525~$\mu$m Si handle substrate, purchased from NanoLN. In general, LNOI wafers can be fabricated using different approaches depending on the target LN device-layer thickness. For thin-film LN below 1~$\mu$m, the ion-slicing process is commonly employed~\cite{ButaudSmart2020a}. This technique relies on ion implantation followed by wafer bonding and thermal exfoliation\cite{levy1998fabrication}, enabling high-quality thin-film LN layers at the expense of increased process complexity and cost. In contrast, for thick LN with thicknesses of several micrometers or more, wafer bonding followed by mechanical lapping and polishing is more commonly used~\cite{ParkPreparation2010}. This approach offers lower cost, simpler processing, and improved wafer availability, while providing the thick LN layer required to realize the high-aspect-ratio fin geometry in this work.

Fabrication of the FinLN devices begins with lithographic patterning and deep Ar ion etching to define high-aspect-ratio LN fin structures (Fig.~\ref{fig:3}). Ar ion etching is a well-established dry etching technique for lithium niobate and has been widely adopted in both integrated LN photonic devices~\cite{ZhuIntegrated2021} and microacoustic LN devices~\cite{WangDesign2015}. Here, the LN etch depth was extended from the typical sub-micrometer regime to approximately 5~$\mu$m to fully define the fin geometry\cite{erturk2023self}. The etch depth was characterized using a KLA-Tencor P-7 stylus profilometer, yielding a step height of approximately 5.3~$\mu$m. This confirms complete removal of LN outside the fin region, with the underlying SiO$_2$ layer exposed and no breaking into the Si substrate. Subsequently, sidewall interdigital transducer electrodes were formed by lithography and e-beam evaporation using glancing-angle deposition (GLAD), depositing 15~nm Ti and 200~nm Al to enable conformal metallization of the fin sidewalls\cite{devitt2024edge,blesin2024bidirectional}.
\begin{figure*}[t]
\centering
\includegraphics[width=\textwidth]{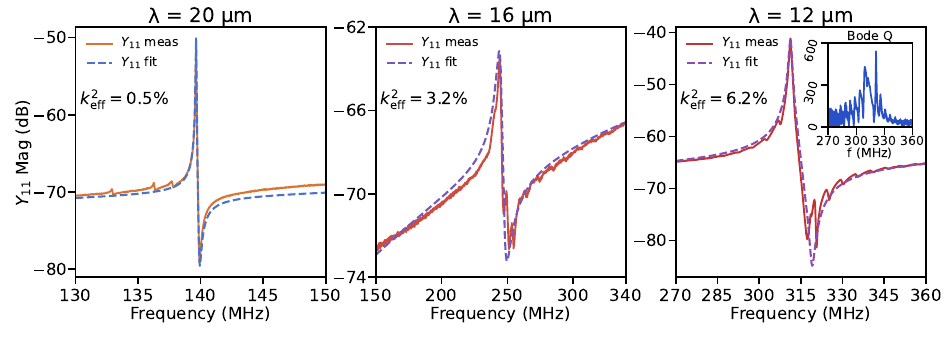}
\caption{
Electrical characterization of FinLN acoustic resonators with different acoustic wavelengths.
Measured admittance magnitude $Y_{11}$ (solid lines) and MBVD model fits (dashed lines) for devices with acoustic wavelengths of $\lambda = 20\,\mu\mathrm{m}$, $16\,\mu\mathrm{m}$, and $12\,\mu\mathrm{m}$, shown from left to right.
The effective electromechanical coupling coefficient $k_{\mathrm{eff}}^{2}$ extracted from the fitting is indicated in each panel.
An inset in the $\lambda = 12\,\mu\mathrm{m}$ panel shows the measured Bode quality factor $Q$ as a function of frequency.
}
\label{fig:5}
\end{figure*}

Fig.~\ref{fig:4} presents representative optical and scanning electron microscopy (SEM) images of the fabricated FinLN acoustic resonators, confirming the successful realization of the high-aspect-ratio fin geometry and sidewall electrode configuration. The top-view images demonstrate uniform fin definition and precise alignment of the interdigital transducer electrodes along the fin axis, consistent with the designed unit-cell geometry and highlighting the compact device footprint. In this device, 50 pairs of IDT electrodes are integrated along the fin in order to achieve a lower series impedance. The fin width and acoustic wavelength of this representative device are designed to be 2~$\mu$m and 12~$\mu$m, respectively, corresponding to the unit-cell parameters used in the simulations. The side-view SEM micrograph further verifies conformal, lithographically defined metallization on the fin sidewalls.

The FinLN resonators were characterized using a Cascade Microtech 11971B probe station equipped with a Keysight N5230A PNA-L network analyzer and Cascade Microtech ACP40 ground--signal--ground (GSG) probes at room temperature under ambient conditions. Devices with a fixed fin width of 2~$\mu$m and acoustic wavelengths of $\lambda = 20~\mu$m, $16~\mu$m, and $12~\mu$m were fabricated and measured to experimentally evaluate the wavelength-dependent trends predicted by the unit-cell simulations. The measured electrical admittance magnitude $Y_{11}$ derived from the probed S-parameters and the corresponding mBVD model fits\cite{larson2000mbvd} are shown in Fig.~\ref{fig:5} for the three devices, arranged from left to right in decreasing acoustic wavelength. From the fitted responses, effective electromechanical coupling coefficients of $k_{\mathrm{eff}}^{2} = 0.5\%$, $3.2\%$, and $6.2\%$ are extracted for $\lambda = 20~\mu$m, $16~\mu$m, and $12~\mu$m, respectively, using Eq.~(\ref{eq:keff})~\cite{LuAccurate2019}. This monotonic increase in $k_{\mathrm{eff}}^{2}$ with decreasing acoustic wavelength is in good agreement with the finite-element simulation results (Fig.\ref{fig:2}).
\begin{equation}
k_{\mathrm{eff}}^{2}
= \frac{\pi^{2}}{8}\frac{C_m}{C_0}
= \frac{\pi^{2}}{8}\frac{f_p^{2}-f_s^{2}}{f_s^{2}},
\label{eq:keff}
\end{equation}
The resonance frequency of the $\lambda = 12~\mu$m device is approximately 310~MHz, consistent with the simulated resonance frequency. Owing to its superior electromechanical performance among the three devices, the quality factor of the $\lambda = 12~\mu$m FinLN resonator was further evaluated using the Bode-$Q$ method, as defined in Eq.~(\ref{eq:bodeQ})~\cite{JinImproved2021}, with the measured result shown in the inset of Fig.~\ref{fig:5}.
\begin{equation}
Q(\omega)
= \omega \left| \frac{\mathrm{d}S_{11}}{\mathrm{d}\omega} \right|
\frac{1}{1-|S_{11}|^{2}},
\label{eq:bodeQ}
\end{equation}
A weak free spectral range (FSR) is observed in both the admittance and Bode-$Q$ responses, which is attributed to longitudinal overtone modes supported by the FinLN structure acting as an acoustic cavity along the fin length. The mode of interest is also a longitudinal mode; the electrode pattern selects one of these modes for excitation, while other longitudinal overtones are only weakly excited.  Consequently, the physically meaningful mechanical quality factor is $Q_m \approx 430$ at the fundamental resonance near 310~MHz. Higher apparent $Q$ values observed near 321~MHz arise from overtone-induced resonances and are therefore not representative of the fundamental FinLN mode.

To quantify the impact of the three-dimensional FinLN geometry and sidewall electrode configuration on electromechanical coupling, a planar reference device was fabricated on the same wafer (Z-cut LN with a 5~$\mu$m LN device layer, a 2~$\mu$m SiO$_2$ layer, and a Si substrate). In this reference device, no fin etching was performed; instead, Ti/Al (15~nm/200~nm) interdigital transducer electrodes were directly patterned on the planar LN surface to form a conventional surface acoustic wave (SAW) resonator, similar to that reported in Ref.~\cite{rubyNovelTemperatureCompensatedSilicon2021}. The detailed device design and measured results of the reference SAW resonator are provided in the Supplementary Material. The planar SAW device shares the same LN film thickness, crystal cut, and IDT orientation as the FinLN resonator, enabling a direct comparison. Using the $\lambda = 12~\mu$m FinLN device as the reference, the measured effective electromechanical coupling coefficient of the planar SAW resonator is approximately $k_{\mathrm{eff}}^{2} \approx 3.4\%$, compared to $k_{\mathrm{eff}}^{2} = 6.2\%$ for the FinLN resonator. This corresponds to an enhancement of approximately 1.8$\times$ in effective electromechanical coupling enabled by the FinLN architecture, highlighting the role of three-dimensional acoustic confinement and sidewall electrode excitation.

The FinLN platform also offers a natural pathway toward integration with photonic devices. The deep-etched lithium niobate geometry employed in this work is compatible with optical waveguide fabrication, and prior studies have shown that deeply etched LN structures can provide reduced fiber-to-chip coupling loss and higher laser damaged threshold\cite{ZhangScalable2023,shi2025hyperfine}. In addition, the surrounding sidewall electrode configuration in the FinLN architecture is expected to enhance acousto-optic interaction strength by improving the spatial overlap between confined acoustic and optical modes. These attributes suggest that the FinLN platform is well suited for integrated acousto-optic devices and hybrid photonic--phononic systems.

In conclusion, we have demonstrated a three-dimensional fin lithium niobate (FinLN) acoustic resonator fabricated from thick lithium niobate using deep ion etching and sidewall electrode integration. The FinLN geometry enables strong three-dimensional acoustic confinement and enhanced electromechanical coupling while maintaining a compact device footprint. The fabricated devices operate near 300~MHz with an effective electromechanical coupling coefficient of 6.2\% and a mechanical quality factor of 430, in good agreement with finite-element simulations. Compared to planar SAW resonators fabricated on the same wafer, the FinLN configuration achieves an approximately 1.8$\times$ enhancement in effective coupling, confirming the advantages of three-dimensional geometry and sidewall excitation. By leveraging thick LNOI wafers produced via wafer bonding and lapping, this approach avoids ion-sliced thin films while remaining compatible with scalable microfabrication, establishing the FinLN architecture as a promising platform for compact, high-performance RF and piezo-optomechanical systems.
\section*{SUPPLEMENTARY MATERIAL}
See supplementary material for the detailed
device design and measured results of the reference SAW resonator.
\begin{acknowledgments}
This material is based upon work supported by the Air Force Office of Scientific Research and the Office of Naval Research under award number FA9550-23-1-0333. Chip fabrication was performed at the Birck Nanotechnology Center at Purdue. Filter and resonator measurements were performed at Seng-Liang Wang Hall at Purdue. The authors thank the Birck Nanotechnology Center staff for their assistance.
\end{acknowledgments}
\section*{DATA AVAILABILITY}
All raw data generated in this study will be released in a Zenodo repository upon publication.

\bibliographystyle{aipnum4-1}
\bibliography{ref}

\end{document}